\documentstyle[12pt]{article}

\begin{document}

\noindent \hspace*{12cm} MAD/PH/749 \\
    \hspace*{12cm} DESY 93--064 \\
    \hspace*{12cm} UdeM-LPN-TH-93-143 \\
    \hspace*{12cm} NUHEP-TH-93-12 \\
\hspace*{12cm} June 1993 \\

\vspace*{1cm}

\centerline{\large{\bf HIGGS BOSONS: INTERMEDIATE MASS RANGE}}

\vspace*{0.4cm}

\centerline{\large{\bf AT e$^+$e$^-$ COLLIDERS}}

\vspace*{1.5cm}

\centerline{\sc V. Barger$^1$, K.~Cheung$^2$, A.~Djouadi$^3$,
B.~A.~Kniehl$^4$ and P.~M.~Zerwas$^5$}

\vspace*{1cm}

\centerline{$^1$ Physics Department, University of Wisconsin, Madison, WI
53706,
USA}

\vspace*{0.2cm}

\centerline{$^2$ Dept. of Physics and Astronomy, Northwestern University,
Evanston, Illinois 60208, USA}

\vspace*{0.2cm}

\centerline{$^3$ Lab. Phys. Nucl., Universit\'e de Montr\'eal, Case 6128A, H3C
3J7 Montr\'eal PQ, Canada}

\vspace*{0.2cm}

\centerline{$^4$ II. Inst.\ f.\ Theor.\ Physik, Universit\"at Hamburg,
D--2000 Hamburg 50, FRG}

\vspace*{0.2cm}

\centerline{$^5$ Deutsches Elektronen--Synchrotron, DESY, D--2000 Hamburg 52,
FRG}

\vspace*{2cm}

\begin{center}
\parbox{14cm}
{\begin{center} ABSTRACT \end{center}
\vspace*{0.2cm}

 We elaborate on the production of the Standard Model Higgs particle at
high-energy $e^+e^-$ colliders through the reaction $e^+e^- \rightarrow  ZH$.
Particular emphasis is put on the intermediate mass range.
In addition to the signal we discuss in detail the background processes.
Angular distributions which are sensitive to the spin and parity of the Higgs
particle are analyzed.}

\end{center}

\newpage

\subsection*{1.~The Physical Basis}

The fundamental particles in the Standard Model (${\cal SM}$), gauge bosons,
leptons and quarks, acquire their masses by means of the Higgs mechanism
\cite{R1}. The masses are generated through the interaction with a non-zero
scalar field in the ground state, inducing the spontaneous breaking of the
electroweak SU(2)$_{\rm I} \times$U(1)$_{\rm Y}$ symmetry down to the
electromagnetic U(1)$_{\rm EM}$ symmetry. To accommodate the well established
electromagnetic and weak phenomena, this mechanism requires the existence of at
least one weak iso-doublet scalar field. The three Goldstone bosons among the
four degrees of freedom are absorbed to build up the longitudinal polarization
states of the massive $W^\pm,Z$ bosons. One degree of freedom is left-over,
corresponding to a real physical particle. The discovery of this Higgs particle
is the {\it experimentum crucis} for the standard formulation of the
electroweak interactions.

 The only unknown parameter in the ${\cal SM}$ Higgs sector is the mass of the
Higgs particle. Even though the value of the Higgs mass cannot be predicted,
interesting constraints can nevertheless be derived from bounds $\Lambda$ on
the energy range in which the model is taken to be valid before perturbation
theory breaks down and new dynamical phenomena could emerge. If the Higgs mass
is less than about 180 to 200 GeV, the Standard Model with weakly interacting
particles can be extended \cite{R2} up the GUT scale $\Lambda_{\rm GUT} \sim
10^{16}$ GeV. This attractive idea is very likely to play a key r\^ole in the
renormalization of the electroweak mixing angle $\sin^2\theta_W$ from the GUT
symmetry value 3/8 down to the experimental value $\sim 0.2$ at low energies. A
lower bound on the Higgs mass follows in the ${\cal SM}$ from the requirement
of vacuum stability for large top quark masses. If, for a given Higgs mass, the
top quark mass increases, radiative corrections drive the scalar potential to
negative values \cite{R3} so that, in turn, for a top mass of $\sim 150$ GeV,
the Higgs mass must exceed $\sim 100$ GeV. We therefore focus in this study on
the important Higgs mass range $M_Z \leq M_H \leq 2M_Z$, generically referred
to as the intermediate Higgs mass range.

 Once the Higgs mass is fixed, the triple and quartic self-couplings of the
${\cal SM}$ Higgs particle are uniquely determined. The scale of the Higgs
couplings to massive gauge bosons and quarks/leptons is set by the masses of
these particles. As a result, the profile of the Higgs boson can be predicted
completely for a given value of the Higgs mass: the decay properties are fixed
and the production mechanisms and rates can be determined.

 The width of the ${\cal SM}$ Higgs particle increases from small values of a
few MeV in the 100 GeV mass range quickly to $\sim 1$ GeV at a mass of $M_H
\sim 200$ GeV. Throughout the intermediate mass range, the width remains so
small that it cannot be resolved experimentally. The dominant decay mode for
masses below $\sim 140$ GeV is the $b\bar{b}$ decay channel [see Fig.~1]. For
higher mass values, the $WW$ and $ZZ$ decays become dominant, one of the vector
bosons being virtual below the threshold for two real bosons. Other decay modes
[$\tau^+ \tau^- , c\bar{c}$ and $gg$] occur at a level of several percent in
the lower part of the intermediate mass range.

 The most comprehensive search for Higgs particles has been carried out in
$Z$ decays at LEP. A lower bound on the Higgs mass of about $M_H \geq 62.5$ GeV
could be established so far \cite{R4}; this limit can be raised to $\sim 65$
GeV by accumulating more statistics. In the second phase of LEP with a total
energy of about 180 GeV, Higgs particles can be searched for up to masses
of about 80 GeV.

 Beyond the LEPII range, new accelerators are needed to search for this
particle and to explore its properties. The multi-TeV $pp$ colliders LHC/SSC
can sweep the entire Higgs mass range up to the ${\cal SM}$ limit of about 800
GeV \cite{R5}. In the upper part of the intermediate range, the lepton channels
$H \rightarrow  ZZ^* \rightarrow  4l^\pm$ will be used, while in the lower part
the rare photon decay $H \rightarrow  \gamma \gamma$ is the sole decay channel,
established so far, which can be exploited to search for this particle. Due to
the overwhelming QCD background, the main decay mode $H \rightarrow  b\bar{b}$
for $M_H \leq 140$ GeV, nor any mode other than the $\gamma \gamma$ channel,
are accessible in hadron colliders. Future $e^+e^-$ colliders, on the other
hand, are the ideal machines to investigate the Higgs sector in the
intermediate mass range since all major decay modes can be explored, with the
Higgs particle produced through several mechanisms \cite{RP}.

 At $e^+e^-$ linear colliders operating in the 300 to 500 GeV energy range,
the main production processes are the $WW/ZZ$ fusion \cite{R8} and the
bremsstrahl \cite{R9} mechanisms. The cross sections for the fusion mechanisms
rise logarithmically with the energy, making these channels dominant at
energies $\geq$ 500 GeV for intermediate mass Higgs bosons. These production
channels have been investigated thoroughly in Ref.~\cite{R7,R9A2,R9A1}. In the
present paper, we analyze in detail the bremsstrahl process
\begin{eqnarray}
e^+e^- \rightarrow  Z^* \rightarrow  Z+H
\end{eqnarray}
throughout the intermediate mass range [see Fig.~2]. The bremsstrahl process is
dominant for moderate values of the ratio $M_H/\sqrt{s}$, but falls off at high
energies according to the scaling law $\sim s^{-1}$. Since this process is
fully constrained, missing mass techniques can be applied which allow to search
for the Higgs particle without any restrictions on the decay modes. [Higgs
events in the $ZZ$ fusion processes $e^+e^- \rightarrow  e^+e^- (ZZ)
\rightarrow  e^+e^- H$ can also be reconstructed completely, yet the production
rate is smaller by an order of magnitude.]

 Once the Higgs boson is found, it will be very important to explore its
properties. This is possible by exploiting the bremsstrahl process (1). The
zero-spin is reflected in the angular distribution \cite{R125}. This
distribution must approach the $\sin^2\theta$ law asymptotically since,
according to the equivalence theorem \cite{ET}, massive gauge bosons act at
high energies like the Goldstone bosons which are absorbed to build up the
longitudinal degrees of freedom. The parity can also be studied by analyzing
angular correlations. Of tantamount importance is the measurement of the Higgs
couplings to gauge bosons and matter particles. The strength of the couplings
to $Z$ and $W$ bosons is reflected in the magnitude of the production cross
section. The relative strength of the couplings to fermions is accessible
through the decay branching ratios. The absolute magnitude is difficult to
measure directly. The direct measurement is possible in the small mass window
where the Higgs decays to $b\bar{b}$ and $WW^*$ compete which each other; else
Higgs bremsstrahlung off top quarks \cite{R9B,ttH} provides an opportunity to
measure the $ttH$ coupling in a limited range of the intermediate mass range.
All these couplings grow with the masses of the source particles, a direct
consequence of the very nature of the mass generation through the Higgs
mechanism. The measurement of these couplings is therefore an important
experimental task in the electroweak sector.

 The paper is organized as follows. In the next section we shall discuss
the production cross section for the bremsstrahl process including radiative
corrections \cite{R10}. Depending on the decay modes, various background
processes, leading to $Zb\bar{b}$ or $ZWW^*,ZZZ^*$ final states, will be
analyzed. We shall also take into account the smearing of the initial state
energy due to conventional photon radiation and beamstrahlung \cite{R11}. In
the third section, we analyze in detail angular distributions and correlations
which allow us to verify the $\cal J^{PC}$
quantum numbers of the Higgs particle. Finally we summarize the methods
which will allow us to measure the couplings
of the Higgs particles to gauge bosons and fermions. No method is known at
this time that could be exploited in practice to measure the self-couplings
of the ${\cal SM}$ Higgs field directly \cite{R12}.

\subsection*{2.~Bremsstrahl Process of Higgs Bosons}

The cross section for the Higgs bremsstrahl process (1) can be written
in a compact form \cite{R7},
\begin{eqnarray}
\sigma (ZH) = \frac{G_F^2M_Z^4}{96 \pi s} \ (a_e^2+v_e^2) \
\beta \ \frac{\beta^2+12M_Z^2/s}{(1-M_Z^2/s)^2}\,,
\end{eqnarray}
where $a_e=-1, v_e=-1+4\sin^2\theta_W$ are the $Z$ charges of the electron, and
$ \beta^2 = [ 1 - (M_H + M_Z )^2$ $/s] [1-(M_H-M_Z)^2/s ]$ is the usual
two-particle phase space factor, proportional to the momentum squared
of the particles in the final
state. The cross section is shown in Fig.~3a for three energy values 300, 400
and 500 GeV as a function of the Higgs mass, and in Fig.~3b for three
representative values of the Higgs mass as a function of the total energy
$\sqrt{ s}$. With $\sigma \sim 200$ fb, a rate of $\sim 4,000$ Higgs particles
in the intermediate mass range is produced at an energy $\sqrt{s}=300$ GeV and
for an integrated luminosity of $\int {\cal L} dt=20$ fb$^{-1}$ within one year
of running. Asymptotically, the cross section scales as $s^{-1}$.

 The radiative corrections {\it sui generis} [without beamstrahlung] are
well under control \cite{R10}. The bulk of these corrections is due to photon
radiation from the incoming electrons and positrons. The size of the radiative
corrections to the total cross section is shown in Fig.~3a/b. The solid curves
include the one-loop weak corrections and the initial-state radiation up to
${\cal O}(\alpha^2)$ with the infrared sensitive parts exponentiated. For small
$M_H$ they are large and positive since photon radiation pulls the $e^+e^-$
system to smaller energies where the cross sections increases. Photon radiation
reduces the cross section, on the other hand, when $M_H$ approaches the
threshold value. The genuine weak corrections have been shown to be small,
reducing the cross section by a few percent [apart from resonating destructive
interference spikes at $M_H=2M_W$ and $2M_Z$].

 For narrow-band beam designs like DLC and TESLA \cite{R14},
beamstrahlung \cite{R11} affects the cross section much less than the standard
initial-state photon radiation. The change of the cross section, relative to
the Born cross section, is exemplified for these design studies in Fig.~4.

 The cleanest channel for isolating the signal from the background is provided
by the $\mu^+\mu^-/e^+e^-$ decay mode of the $Z$ boson. Depending on the mass
of the Higgs boson, either $H \rightarrow  b\bar{b}$ decays dominate for $M_H
\leq 140$ GeV, or $H \rightarrow  WW^*$ decays for $M_H \geq 140$ GeV. We shall
assume a perfect $\mu$-vertex detector with 100\% $B$ detection efficiency,
which is not far from experimental reality \cite{R16}. $Z \rightarrow  \mu^+
\mu^-$ can be tagged by allowing the invariant $\mu^+ \mu^-$ mass to vary
within $|M_{\mu^+ \mu^-}-M_Z| \leq 6$ GeV. For the cases $H\rightarrow
VV^*\;(V=W,Z)$, we tag the hadronic decays of the on-shell vector boson by
requiring that $|M_{jj} - M_V|< M_V/10$. While the signal events prefer the
central region of the detector, as discussed later in detail, the background
reactions $e^+e^- \rightarrow  Z+X$ are in general strongly peaked in the
forward direction due to the $t$-channel exchange of electrons \cite{R125}. A
cut $|\cos \theta_Z| <0.6$ therefore improves the signal-to-noise ratio
significantly. For the study of $H\rightarrow b\bar b$  the dominant
backgrounds come from $e^+e^-\rightarrow ZZ^*,\,Z\gamma^*,\,\gamma^*\gamma^*
\rightarrow (\mu^+\mu^-)(b\bar b)$, as well as $e^+e^-\rightarrow  t\bar t
\rightarrow  (b\mu^+\nu) (\bar b\mu^- \bar \nu)$. The other backgrounds like
$e^+e^-\rightarrow Zb\bar b$ are suppressed by one more factor of $\alpha_{\rm
W}$. On the other hand, for $H\rightarrow WW^*$  the dominant background comes
from the continuum production of $e^+e^-\rightarrow ZWW^*\rightarrow
(\mu^+\mu^-)+X$.  The backgrounds for the channel $H\rightarrow ZZ^*$ are more
complicated. Besides the continuum production of $e^+e^-\rightarrow ZZZ^*$,
there are also combinatorial ambiguities in the signal process from the chain
$e^+e^- \rightarrow ZH \rightarrow (jj) (ZZ^*)\rightarrow (jj)(\mu^+\mu^-jj)$,
which will not contribute to the Higgs peak in the missing mass spectrum but
rather broaden the continuum (see Fig.~5c and 6c). Without loss of a large
fraction of the signal events, we can assume that all the jets are very well
separated, so that the QCD cross section $\sigma(e^+e^-\rightarrow q\bar qggZ)$
is of ${\cal O}(\alpha_{\rm W}^3\alpha_{\rm s}^2)$, i.e., suppressed by a
factor of $\alpha_{\rm W}\alpha_{\rm s}^2$ compared to the signal. The main
background reactions are summarized in Table 1.

\vspace*{0.4cm}

\begin{tabular}{|c|c|c|c|} \hline
& & & \\
channel & $ H  \rightarrow  b\bar{b}$ & $H \rightarrow  WW^*$ & $ H\rightarrow
ZZ$ \\
& $M_H < 160$~GeV &  $M_H >110$~GeV & $M_H > 180$~GeV \\
& & & \\ \hline
& & & \\
signal & $ e^+e^- \rightarrow  (\mu^+ \mu^-) (b \bar{b}) $
       & $ e^+e^- \rightarrow  (\mu^+ \mu^-) (W     W^*) $
       & $ e^+e^- \rightarrow  (\mu^+ \mu^-) (Z     Z) $ \\
& & & \\ \hline
& & & \\
bkgd & $ e^+e^- \rightarrow  ZZ, Z \gamma^*, \gamma^* \gamma^* $
     & $ e^+e^- \rightarrow  ZWW^* $
     & $  e^+e^- \rightarrow  ZZZ $ \\
& $e^+e^-\rightarrow  t\bar t \rightarrow  b\bar b \mu^+\mu^-\nu\bar \nu$
& &  $[e^+e^- \rightarrow  ZH \rightarrow  (jj)(\mu^+\mu^- jj)]$ \\
& & & \\ \hline
\end{tabular}

\vspace*{0.3cm}

{\bf Table 1}. Signal and background processes to Higgs production via
bremsstrahlung, \\ \hspace*{2.2cm} $e^+e^- \rightarrow  HZ$, for various mass
ranges.

\vspace*{0.3cm}

 The distributions with respect to the missing mass,
$M_{\rm recoil}^2=\left[ (p_{e^-}+p_{e^+})-(p_{\mu^-}+p_{\mu^+}) \right]^2$,
are shown in Figs.~5a--c for the three channels defined in Table~1
assuming $\sqrt s=500$~GeV.
In each channel we consider three representative values of $M_H$:
$M_H=100,125,150$~GeV in the $H\to b\bar b$ case,
$M_H=120,150,180$~GeV in the $H\to WW^*$ case, and
$M_H=150,200,250$~GeV in the $H\to ZZ^*$ case.
In Figs.~6a--c, we show the corresponding results for $\sqrt s=300$~GeV.
Signals and backgrounds have been added, the cuts specified above have been
applied, and initial-state bremsstrahlung and beamstrahlung corrections
have been taken into account assuming the DLC narrow-band design.
For this design, the smearing due to beamstrahlung is small at
$\sqrt s=500$~GeV and almost negligible at $\sqrt s=300$~GeV.
Without the energy loss connected with these initial-state radiative effects
the signals would appear as a sharp peak at $M_{\rm recoil}=M_H$.
Throughout our analysis we have identified jets with the partons in which they
originate.

 In Figs.~5a and 6a, the $ZH\rightarrow  (\mu^+\mu^-)(b\bar b)$ signal
is shown with its background added. The $M_{\rm recoil}$ distribution starts
slightly below $M_H$, rises steeply to the peak at $M_H$, and tails off towards
larger values of $M_{\rm recoil}$. For $M_H\le 100$~GeV, the suppression of the
$ZZ$ background by $\mu$-vertexing is essential. For larger $M_H$, the Higgs
and $Z$ peaks are clearly separated. Figures 5b and 6b show the $M_{\rm
recoil}$ spectrum for the $ZH\rightarrow  (\mu^+\mu^-) WW^*$ signal with the
continuum background added. For $M_H=120$~GeV, the Higgs peak is low due to the
small branching fraction of $H\rightarrow  WW^*$. Once $M_H\ge 140$~GeV,
BR($H\rightarrow  WW^*$) increases, however, quite fast with $M_H$ [see
Fig.~1b] and the Higgs peaks are very prominent in the curves for $M_H=150$ and
180~GeV. All these curves will eventually merge together at the high end of the
$M_{\rm recoil}$ spectrum because only the continuum background of
$e^+e^-\rightarrow  ZWW^*$ contributes there. Figures 5c and 6c exhibit
essentially the same features as Figs.~5b and 6b, except that the curves in
these figures will not merge at high $M_{\rm recoil}$. This is due to the
presence of the combinatorial ``background'' from $e^+e^-\rightarrow  ZH
\rightarrow (jj)(\mu^+\mu^- jj)$, which will not contribute to the Higgs peak
but to the continuum; in particular, the increase of the background at large
$M_{\rm recoil}$ for the 200~GeV case is due to combinatorics. In conclusion,
$M_H$ can be easily determined from the $M_{\rm recoil}$ spectrum.

\subsection*{3.~The ${\cal J^{PC}}=0^{++}$ Quantum Numbers of the Higgs
Boson}

The scalar character of the Higgs particle can be tested at $e^+e^-$ colliders
in several ways. The angular distribution of the $e^+e^- \rightarrow  ZH$ final
state depends on the spin and parity of the Higgs particle. The same is true of
the angular correlations in the Higgs decay to fermion and gauge boson pairs.
The observation of the decay or fusion processes $H \rightleftharpoons \gamma
\gamma$ would rule out spin =1 directly by Yang's theorem and fixes the charge
conjugation to be positive ${\cal C}=+$. This follows also from the decay and
fusion processes $H \rightleftharpoons ZZ$, as well as from the observation of
the bremsstrahl process $e^+e^-\rightarrow Z^*\rightarrow ZH$.

\subsection*{3.1.~Production Process}

Since the production of the final state $e^+e^- \rightarrow  Z^* \rightarrow
ZH$ is mediated by a virtual $Z$ boson [transversally polarized along the
$e^\pm$ beam axis], the production amplitude could be a monomial in
$\cos/\sin\theta$. In the $\cal SM$, however, the $ZZH$ coupling,
\begin{eqnarray}
{\cal L}(ZZH)&=&\left(\sqrt2G_F\right)^{1/2}M_Z^2HZ^\mu Z_\mu \nonumber \\
&\sim&G_F^{1/2}HF^{\mu\nu}F_{\mu\nu}\,,
\end{eqnarray}
is an S-wave coupling
$\sim \vec\epsilon_1\cdot\vec\epsilon_2$ in the laboratory frame,
linear in $\sin \theta$ and even under parity and charge conjugation,
corresponding to the $0^{++}$ assignment of the Higgs quantum numbers.  The
explicit form of the angular distribution is given by
\begin{eqnarray}
\frac{ {\rm d}\sigma (ZH)} { {\rm d} \cos \theta} \sim \beta^2
\sin^2\theta +8M_Z^2/s\,.
\end{eqnarray}
For high energies, the $Z$ boson is produced in a state of longitudinal
polarization,
\begin{eqnarray}
\frac{\sigma_L} {\sigma_L +\sigma_T} = 1- \frac{8M_Z^2}{12M_Z^2+\beta^2s}\,,
\end{eqnarray}
so that---in accordance with the equivalence theorem \cite{ET}---the production
amplitude becomes equal to the amplitude
${\cal M}(e^+e^- \rightarrow  \Phi^0H)$, with $\Phi^0$
being the neutral Goldstone boson which is absorbed to build up the
longitudinal degrees of freedom of the electroweak bosons. The angular
distribution therefore approaches the spin-zero distribution asymptotically,
\begin{eqnarray}
\frac{1} {\sigma} \frac{ {\rm d}\sigma(ZH)} { {\rm d} \cos \theta}
\ \rightarrow  \ \frac{3}{4} \sin^2\theta\,.
\end{eqnarray}

 Even though the nature of the Higgs phenomenon requires the Higgs field in
the ${\cal SM}$ to be necessarily a scalar ${\cal J^{PC}}=0^{++}$ field, it is
nevertheless interesting to confront the predictions in (4,5) with the
production
\begin{eqnarray}
e^+e^- \rightarrow  Z^* \rightarrow  ZA
\end{eqnarray}
of a pseudoscalar state $A(0^{-+})$ in order to underline the uniqueness of the
$\cal SM$ prediction. The pseudoscalar case is realized in two-doublet Higgs
models, in which the $ZZA$ couplings are induced by higher-order loop effects
\cite{HHG}. The effective point-like coupling,
\begin{eqnarray}
{\cal L}(ZZA)={\eta\over4}\left(\sqrt2G_F\right)^{1/2}M_Z^2A
Z^{\mu\nu}\widetilde Z_{\mu\nu}\,,
\end{eqnarray}
with $\eta$ being a dimensionless factor and $\widetilde
Z^{\mu\nu}=\epsilon^{\mu\nu\rho\sigma}Z_{\rho\sigma}$, is a P-wave coupling,
odd under parity and even under charge conjugation. It reduces to
$(\vec{\epsilon}_1 \times \vec{\epsilon}_2)\cdot(\vec{p}_1 - \vec{p}_2)$ in the
laboratory frame. Since the $Z$ spins are coupled to a vector, the angular
distribution is again a binomial in $\sin \theta$,
\begin{eqnarray}
\frac{ {\rm d}\sigma (ZA)} { {\rm d} \cos \theta} \sim 1- \frac{1}{2}
\sin^2\theta\,,
\end{eqnarray}
independent of the energy. The $Z$ boson in the final state is purely
transversally polarized so that the cross section need not be $\sim
\sin^2\theta$ in this case. The total cross section is given by
\begin{eqnarray}
\sigma (ZA) = \eta^2\frac{G_F^2M_Z^6}{48\pi M_H^4} \ (a_e^2+v_e^2) \
\ \frac{\beta^3}{(1-M_Z^2/s)^2}\,,
\end{eqnarray}
where the momentum dependence $\sim \beta^3$ is
characteristically different from the $ZH$ production cross section
near threshold.

 The angular correlations specific to the S-wave $0^{++}$ Higgs production
in $e^+e^- \rightarrow  ZH$ can directly be confronted experimentally
with the process $e^+e^- \rightarrow  ZZ$. This process has an angular momentum
structure that is distinctly different from the Higgs process. Mediated by
electron exchange in the $t$-channel, the amplitude is built-up by many partial
waves, peaking in the forward/backward direction. The two distributions are
compared with each other in Fig.~7. This figure demonstrates the specific
character of the Higgs production process, which is not lost if experimental
acceptance cuts and smearing effects are taken into account \cite{R9A2}.

 Since the longitudinal wave function of a vector boson grows with the
energy of the particle---in contrast to the energy independent transverse
wave function---the $Z$ boson in the S-wave Higgs production process (1)
must asymptotically be polarized longitudinally, Fig.~8. By contrast, the
$Z$ bosons from $ZA$ associated production or $ZZ$ pair production
are transversally polarized [at high energies in the second case].
This pattern can be checked experimentally. While the distribution
of the light fermions in the $Z \rightarrow  f\bar{f}$ rest frame with respect
to the flight direction of the $Z$ [see Fig.~9a] is given by $\sin^2\theta_*$
for longitudinally polarized $Z$ bosons, it behaves as $(1\pm \cos\theta_*)^2$
for transversally polarized states, after averaging over the azimuthal angles.
Including the azimuthal angles, the final angular correlations may be written
for $e^+e^- \rightarrow  ZH$ $[Z\rightarrow  f \bar{f}]$ as
\begin{eqnarray}
\frac{ {\rm d}\sigma (ZH)} { {\rm d} c_\theta {\rm d} c_{\theta_*}
{\rm d} \phi_*} &\sim & s^2_\theta s^2_{\theta_*} -\frac{1}{2 \gamma} s_{2
\theta} s_{2\theta_*} c_{\phi_*} + \frac{1}{2\gamma^2} [(1+c^2_\theta)(1
+c^2_{\theta_*})+s^2_\theta s^2_{\theta_*} c_{2\phi_*} ] \nonumber \\
&&-\ \frac{2v_ea_e}{v_e^2+a_e^2}\,\frac{2v_fa_f}{v_f^2+a_f^2}\,\frac{2}{\gamma}
\left[ s_\theta s_{\theta_*} c_{\phi_*} -\frac{1}{\gamma} c_\theta c_{\theta_*}
   \right]\,,
\end{eqnarray}
where $s_\theta= \sin \theta$, $etc$. As before, $\theta$ is the polar $Z$
angle in the laboratory frame, $\theta_*$ the polar fermion angle in the $Z$
rest frame and $\phi_*$ the corresponding azimuthal angle with respect to the
$e^\pm ZH$ production plane. After integrating out the polar angles $\theta$
and $\theta_*$, we find the familiar $\cos \phi_*$ and $\cos 2 \phi_*$
dependence associated with ${\cal P}$-odd and even amplitudes, respectively,
\begin{eqnarray}
\frac{ {\rm d}\sigma (ZH)} { {\rm d} \phi_*} \sim 1+a_1 \cos \phi_*
+ a_2 \cos 2\phi_*\,,
\end{eqnarray}
with
\begin{eqnarray}
a_1= -\frac{9\pi^2}{32}\,\frac{\gamma}{\gamma^2+2}\,
\frac{2v_ea_e}{v_e^2+a_e^2}
\frac{2v_fa_f}{v_f^2+a_f^2}\,,\qquad
a_2= \frac{1}{2}\,\frac{1}{\gamma^2+2}\,.
\end{eqnarray}
The azimuthal angular dependence disappears for high energies $\sim 1/\gamma$
as a result of the dominating longitudinal polarization of the $Z$ boson.

 Note again the characteristic difference to the pseudoscalar $0^{-+}$ case
$e^+e^- \rightarrow  ZA$ $[Z \rightarrow  f\bar{f}]$,
\begin{eqnarray}
\frac{ {\rm d} \sigma (ZA)} { {\rm d} c_\theta {\rm d} c_{\theta_*}
{\rm d} \phi_*} &\sim & 1+ c^2_\theta c^2_{\theta_*} -\frac{1}{2}
s^2_\theta s^2_{\theta_*} -\frac{1}{2} s^2_\theta s^2_{\theta_* }
c_{2\phi_*} \nonumber \\
& &+ 2 \frac{2v_ea_e}{v_e^2+a_e^2}\,\frac{2v_fa_f}{v_f^2+a_f^2}c_\theta
c_{\theta_*}\,.
 \end{eqnarray}
This time the azimuthal dependence is ${\cal P}$-even and independent of the
energy in contrast to the $0^{++}$ case; after integrating out the polar
$\theta, \theta_*$ angles,

\begin{eqnarray}
\frac{ {\rm d}\sigma (ZA)} { {\rm d} \phi_*} \sim 1 -\frac{1}{4}
\cos 2\phi_*\,.
\end{eqnarray}

We can thus conclude that the angular analysis of the Higgs production in
$e^+e^- \rightarrow  Z^* \rightarrow  ZH$ $[Z \rightarrow  f \bar{f}]$ allows
stringent tests of the ${\cal J^{PC}}=0^{++}$ quantum numbers of the Higgs
boson. This is a direct consequence of the $0^{++}$ coupling
$\epsilon_1\cdot\epsilon_2$ of the $ZZH$ vertex in the production amplitude.

\subsection*{3.2.~Decay Processes}

The zero-spin of the Higgs particle can be checked directly through the lack of
any angular correlation between the initial and final state particles
\cite{R9A2}. In the following discussion we shall concentrate on Higgs boson
decays to vector bosons $V=W,Z$, which will provide us with signatures similar
to those studied in the previous section, including the discrimination between
$0^{++}$ and $0^{-+}$ decays. The analysis applies in a straightforward way to
Higgs particles in the $e^+e^-$ environment. Background problems require a more
sophisticated discussion for Higgs particles produced at the SSC and LHC. Since
some of the material on this method had been worked out before \cite{R18}, we
focus on novel points which have not been elaborated in detail so far.

 Above the $H \rightarrow  WW$ and $ZZ$ decay thresholds, the partial width
into massive gauge boson pairs may be written as \cite{R19}
\begin{eqnarray}
\Gamma (H \rightarrow  VV) = \delta_V \frac{\sqrt{2}G_F}{32 \pi} M_H^3
(1-4x+12x^2)
\beta\,,
\end{eqnarray}
where $x=M_V^2/M_H^2$, $\beta=\sqrt{1-4x}\,$ and $\delta_V=2(1)$ for $V=W(Z)$.
For large Higgs masses, the vector bosons are longitudinally polarized
[see Fig.~10],
\begin{eqnarray}
\frac{\Gamma_L}{\Gamma_L+\Gamma_T} =  \frac{1-4x+4x^2}{1-4x+12x^2}\,,
\end{eqnarray}
while the $L,T$ polarization states are democratically populated near the
threshold, at $x=1/4$.
 The $HZZ$ coupling $\epsilon_1\cdot\epsilon_2$ is the same S-wave
coupling which we have discussed earlier. Since the longitudinal wave functions
are linear in the energy, the width grows as the third power of the Higgs mass.
The electroweak radiative corrections \cite{R20} are positive and amount to
a few percent above the threshold.

 Below the threshold for two real bosons, the Higgs particle can decay into
real and virtual $VV^*$ pairs, primarily $WW^*$ pairs above $M_H \sim 110$ GeV.
The partial decay width, $W$ charges summed over, is given \cite{R21} by
\begin{eqnarray}
\Gamma (H \rightarrow  VV^*) = \frac{3 G_F^2 M_V^4}{16 \pi^3} M_H R(x)
\delta_V'\,,
\end{eqnarray}
with $\delta'_W=1$ and $\delta_Z' =7/12-10\sin^2\theta_W/9+40\sin^4\theta_W/27$
and
\begin{eqnarray}
R(x)=\frac{3(1-8x+20x^2)}{(4x-1)^{1/2}} \arccos \left( \frac{3x-1}{2x^{3/2}}
\right) -\frac{1-x}{2x} (2-13x+47x^2) - \frac{3}{2}(1-6x+4x^2)
\log x\,. \hspace*{0.3cm}
\end{eqnarray}

 The invariant mass ($M_*$) spectrum of the off-shell vector
boson peaks close to the kinematical maximum corresponding to
zero-momentum of the on- and off-shell vector bosons in the final state
[see Fig.~11],
\begin{eqnarray}
\frac{{\rm d} \Gamma}{{\rm d}M_*^2} = \frac{3G_F^2 M_V^4} {16 \pi^3M_H}
\delta_V' \ \frac{\beta (M_H^4\beta^2 + 12M_V^2 M_*^2)}{(M_*^2-M_V^2)^2+M_V^2
\Gamma_V^2}\,,
\end{eqnarray}
where
$\beta^2=\left[1-(M_V+M_*)^2/M_H^2\right]\left[1-(M_V-M_*)^2/M_H^2\right]$.
Since both $V$ and $V^*$ preferentially have small momenta, the transverse and
longitudinal polarization states are populated with almost equal probabilities,
\begin{eqnarray}
\frac{ {\rm d} \Gamma_L / {\rm d}M_*^2} { {\rm d}\Gamma / {\rm d}M_*^2}
=1 - \frac{8M_V^2M_*^2}{M_H^4\beta^2+12 M_V^2M_*^2}\,.
\end{eqnarray}
Neglecting the widths of the vector bosons, $\Gamma_V$,
we find after summing over all $M_*$ values [see Fig.~10]
\begin{eqnarray}
\frac{\Gamma_L}{\Gamma_L+\Gamma_T}=\frac{R_L(M_V^2/M_H^2)}{R(M_V^2/M_H^2)}\,,
\end{eqnarray}
where $R$ is given in eq.~(19) and
\begin{eqnarray}
R_L(x)=\frac{3-16x+20x^2} {(4x-1)^{1/2}} \arccos \left( \frac{3x-1}{2x^{3/2}}
\right) -\frac{1-x}{2x} (2-13x+15x^2) - \frac{1}{2}(3-10x+4x^2) \log x\,.
\hspace*{0.3cm}
\end{eqnarray}

 The angular distributions of the fermions in the $0^{++}$ decay process
\begin{eqnarray}
H \rightarrow  V \ V^* \rightarrow  (f_1 \bar{f}_2) \ (f_3 \bar{f}_4) \qquad
[V=W,Z]
\end{eqnarray}
are quite similar to the rules found for the production process.
Denoting
polar and azimuthal angles of the fermions $f_1,f_3$ in the rest frames of
the vector bosons by $(\theta_1,0)$ and $(\theta_3,\phi_3)$, respectively,
[see Fig.~9b] the angular distributions for the scalar case are given by
\begin{eqnarray}
\frac{ {\rm d}\Gamma (H \rightarrow  VV^*)} { {\rm d} c_{\theta_1} {\rm
d}c_{\theta_3}
{\rm d} \phi_3} &\sim & \ \ s^2_{\theta_1} s^2_{\theta_3} +\frac{1}{2 \gamma_1
\gamma_3(1+\beta_1 \beta_3)} s_{2 \theta_1} s_{2\theta_3} c_{\phi_3}
\nonumber \\
& &+{1\over2\gamma_1^2\gamma_3^2(1+\beta_1\beta_3)^2}
\left[\left(1+c_{\theta_1}^2\right)\left(1+c_{\theta_3}^2\right)
+s_{\theta_1}^2s_{\theta_3}^2c_{2\phi_3}\right]
\nonumber \\
& &-\frac{2v_1 a_1}{v_1^2+a_1^2}\,\frac{2v_3a_3}{v_3^2+a_3^2}\,
\frac{2}{\gamma_1\gamma_3(1+\beta_1 \beta_3)}
\left[ s_{\theta_1} s_{\theta_3} c_{\phi_3}
+\frac{1}{\gamma_1\gamma_3(1+\beta_1 \beta_3)}
c_{\theta_1}c_{\theta_3}\right]\,.\qquad
\end{eqnarray}
For $V=W$, the charges are $v_i=a_i=1$; for $V=Z$, $v_i=2I_{3i}-4e_i
\sin^2\theta_W$ and $a_i=2I_{3i}$. $\beta_i, \gamma_i$ are the velocities and
$\gamma$ factors of the [on/off-shell] vector bosons.
As expected, the dependence on the azimuthal angle between the decay planes
disappears for large Higgs masses, a consequence of the asymptotic longitudinal
$V$ polarization. After integrating out the polar angles, we are left with
\begin{eqnarray}
\frac{ {\rm d}\Gamma (H \rightarrow  VV^*)} { {\rm d} \phi_3} \sim 1+a_1 \cos
\phi_3
+ a_2 \cos 2\phi_3\,,
\end{eqnarray}
where
\begin{eqnarray}
a_1 &=& -\frac{9 \pi^2}{32}\,\frac{\gamma_1 \gamma_3 (1+\beta_1 \beta_3)}
{\gamma_1^2\gamma_3^2 (1+\beta_1 \beta_3)^2+2}\,\frac{2v_1a_1}{v_1^2+a_1^2}
\,\frac{2v_3a_3}{v_3^2+a_3^2}\,, \nonumber \\
a_2& =&  \frac{1}{2}\,\frac{1}
{\gamma_1^2 \gamma_3^2 (1+\beta_1 \beta_3)^2+2}\,.
\end{eqnarray}
The coefficient $a_1$ measures the ${\cal P}$-odd amplitude.

 Both the azimuthal angular correlations as well as the distributions of the
polar angles are sensitive to the spin-parity assignments of the Higgs
particle, which affect strongly the
populations of the $W,Z$ polarization states.
We will study this point by confronting again the scalar $H(0^{++})$ decay
with the pseudoscalar $A(0^{-+})$ decay distributions,
\begin{eqnarray}
A \rightarrow  V \ V^* \rightarrow  (f_1 \bar{f}_2) \ (f_3 \bar{f}_4) \qquad
[V=W,Z]\,.
\end{eqnarray}
For the $AV_{\mu \nu}\widetilde{V}^{\mu \nu}$ coupling defined before,
the angular distributions is given by
\begin{eqnarray}
\frac{ {\rm d}\Gamma (A \rightarrow  VV^*)} { {\rm d} c_{\theta_1} {\rm
d}c_{\theta_3}
{\rm d} \phi_3} & \sim & 1+ c^2_{\theta_1} c^2_{\theta_3} -\frac{1}{2}
s^2_{\theta_1} s^2_{\theta_3} -\frac{1}{2} s^2_{\theta_1} s^2_{\theta_3}
c_{2\phi_3} \nonumber \\
& &-2 \frac{2v_1 a_1}{v_1^2+a_1^2}\,\frac{2v_3a_3}{v_3^2+a_3^2}
c_{\theta_1} c_{\theta_3}\,.
\end{eqnarray}
The normalization follows from the total and differential widths,
\begin{eqnarray}
\Gamma (A \rightarrow  VV^*) = \frac{3 G_F^2 M_V^6}{8 \pi^3 M_A} \delta_V'
\eta^2
A \left( \frac{M_V^2}{M_A^2} \right)
\end{eqnarray}
with
\begin{eqnarray}
A(x)=(1-7x)(4x-1)^{1/2} \arccos \left( \frac{3x-1}{2x^{3/2}}
\right) -\frac{1-x}{6}(17-64x-x^2) + \frac{1}{2}(1-9x+6x^2) \log x\,,
\hspace*{0.3cm}\nonumber
\end{eqnarray}
and
\begin{eqnarray}
\frac{{\rm d} \Gamma (A \rightarrow  VV^*)}{{\rm d}M_*^2} = \frac{3 G_F^2
M_V^6}{8 \pi^3
M_A} \delta_V' \eta^2 \frac{M_*^2 \beta^3}{(M_*^2-M_V^2)^2+ M_V^2 \Gamma_V^2}
\,,
\end{eqnarray}
where $M_*$ denotes the mass of the off-shell vector boson. The $M_*$ spectra
for $H\rightarrow  Z^*Z$ and $A\rightarrow  Z^*Z$ are shown in Fig.~11 assuming
$M_H=M_A=150$~GeV. The mass and momentum dependence of the width are determined
by the P-wave decay characteristics and the transverse polarization of the
vector bosons. In the same way, the angular distributions [which are
independent of the masses] reflect the transverse $W,Z$ polarizations.

 For Higgs masses below 140 GeV, the dominant decay mode are $b\bar{b}$
decays. A fraction of a few percent decay into $\tau^+ \tau^-$ final states.
For moderate to large tg$\beta$ values, these are also the main decay modes for
scalar and pseudoscalar Higgs particles in the minimal supersymmetric
extension of the Standard Model ($\cal MSSM$).
While the zero-spin can again be checked experimentally
by the lack of any correlations between the final and initial state particles,
it is quite difficult to verify the $\pm$ parity of the states in the decay
distributions. This information must be extracted from correlations between the
$b$ and $\bar{b}$, $\tau^-$ and $\tau^+$ spin components in the
planes transverse to the $b{\bar b}$ and $\tau^+ \tau^-$ axes,
\[
{\rm scalar/pseudoscalar} \ : \ C=1- s_{||}^fs_{||}^{\bar{f}} \pm
s_{\perp}^fs_{\perp}^{\bar{f}}
\]
The spin correlations are reflected in correlations between the $f$ and
$\bar{f}$ decay products \cite{R22}. Depolarization effects through $b,\bar{b}$
fragmentation into pseudoscalar $B$ mesons \cite{R23} introduce systematic
uncertainties into this channel which are very hard to control, so that the
rare but clean $\tau \tau$ final states are the proper instrument to study this
problem.

 At $e^+e^-$ colliders, the parity of the Higgs bosons in the ${\cal MSSM}$ can
be studied in the following way.
\begin{description}
\item[(i)] \ \ Since the scalar ${\cal MSSM}$ Higgs particles $h,H$ couple to
vector
bosons directly, the positive parity can be checked by analyzing the $Z$
final states
in $e^+e^- \rightarrow  Z^* \rightarrow  ZH\ [Z \rightarrow  f \bar{f}]$ as
discussed above in great detail.
This method is equivalent to the analysis of the $h,H \rightarrow  VV$ decays.
\item[(ii)] \ The fusion of Higgs particles by linearly polarized
photon beams depends on the angle between the polarization vectors \cite{R24}.
For scalar particles the production amplitude $\sim \vec{\epsilon}_1\cdot
\vec{\epsilon}_2$ is non-zero only for parallel vectors while pseudoscalar
particles with amplitudes $\sim\vec{\epsilon}_1 \times \vec{\epsilon}_2$
require perpendicular polarization vectors.
For typical experimental set-ups for Compton back-scattering of laser light
\cite{RA1}, the maximum degree of linear polarization of the generated
hard photon beams is less than about 30\% so that the efficiency for two
polarized beams is reduced to less than 10\% \cite{RA2}.
This method therefore requires high luminosities, and, moreover, a careful
analysis of background rejection due to the enormous number of $b\bar b$
pairs produced in $\gamma\gamma$ collisions \cite{RA3}.
\end{description}

\subsection*{4.~Summary}

In this report we have described the production of Higgs particles in the
intermediate mass range and the analysis of their ${\cal J^{PC}}$ assignment.
Even though we have focused on $e^+e^-$ linear colliders, many elements of our
study on Higgs decays are also relevant to Higgs particles eventually produced
at the proton colliders SSC and LHC.

It has been shown that the search for ${\cal SM}$ Higgs particles in the
intermediate mass range is easy at $e^+e^-$ colliders since the signal sticks
out of the background processes very clearly. $\mu$-vertexing of the $b$ quarks
will be essential for Higgs bosons in the $Z$-mass range. The spin and parity
assignments can, on one hand, be checked easily in Higgs decays to vector boson
pairs. Angular correlations among the lepton/quark decay products reflect their
polarization states, primarily longitudinal or democratic for large or moderate
Higgs masses, respectively, if ${\cal J^{PC}}=0^{++}$, and transverse if ${\cal
J^{PC}}=0^{-+}$. This information can also be extracted, on the other hand,
from the angular behavior of the $e^+e^- \rightarrow  ZH$ production process.
The parity analysis of pseudoscalar Higgs particles, in supersymmetric
extensions of the Standard Model for instance, requires difficult measurements
of the fermionic decay states, or polarization analyses in the case of $\gamma
\gamma$ fusion.

 Besides the spin-parity assignments, the essential nature of the ${\cal SM}$
Higgs particle must be established by demonstrating that couplings to other
fundamental particles grow with their masses. Even though we did not analyze
these couplings in detail, a few remarks ought to be added to round off the
discussion \cite{R7}. The Higgs couplings to massive gauge bosons can directly
be determined from the measurement of the production cross sections: the $HZZ$
coupling in the bremsstrahl and in the $ZZ$ fusion processes; the $HWW$
coupling in the $WW$ fusion process.   For sufficiently large Higgs masses
above $\sim$ 250 GeV, these couplings can also be determined experimentally
from the decay widths $H \rightarrow  ZZ, WW$. Higgs couplings to fermions are
not easy to measure directly. For Higgs bosons in the intermediate mass range
where the decays into $b\bar{b},c \bar{c}$ and $\tau^+ \tau^-$ are important,
the decay width is so narrow that it cannot be resolved experimentally.
Nevertheless, the branching ratios into $\tau$ leptons and charm quarks reveal
the couplings of these fermions relative to the coupling of the $b$ quarks into
which the Higgs boson decays predominantly. In the upper part of the
intermediate mass range but below the threshold for real $WW$ decays, the
branching ratio BR($H \rightarrow  WW^*$) is sizeable and can be determined
experimentally \cite{R25}. In this case, the absolute values of the $b$ and
eventually of the $c , \tau$ couplings can be derived once the $HZZ/HWW$
couplings are fixed by the production cross sections. The decays $H \rightarrow
gg$ and $\gamma \gamma , Z \gamma $ and the fusion processes $gg,\gamma \gamma
\rightarrow  H$ are mediated by loop diagrams and they are proportional to the
couplings of the Higgs boson to heavy particles.  The number of heavy particles
can be counted in these processes if their masses are generated by the Higgs
mechanism and their couplings to the Higgs particles grow with the mass. [This
is not the case for heavy supersymmetric particles which decouple
asymptotically from the $HVV$ vertices.] In the Standard Model only the top
quark contributes to the $Hgg$ vertex so that the $Htt$ coupling can be
measured in the gluonic decays of the Higgs particle [and, in the same way,
through the cross section for the gluon fusion $gg \rightarrow  H$ at hadron
colliders]. In the case of $H \rightarrow  \gamma \gamma, Z \gamma$ additional
contributions to the decay amplitudes come from $W$ loops. A direct way to
determine the Yukawa coupling of the intermediate mass Higgs boson to the top
quark in the range $m_H \le 120$ GeV are  provided by the bremsstrahl process
$e^+e^-\rightarrow   t\overline{t} H$ in high energy $e^+e^-$ colliders
\cite{R9B} and $\gamma\gamma\rightarrow  t\bar tH$ in photon-photon colliders
\cite{ttH}. For large Higgs masses above the $t\bar{t}$ threshold,  the decay
channel $H \rightarrow  t\bar{t}$ increases the cross section of $e^+e^-
\rightarrow  t\bar{t}Z$ through the reaction $e^+e^- \rightarrow
ZH(\rightarrow  t\bar{t}\,)$ \cite{R9C}; without the Higgs decay this final
state is produced mainly through virtual $\gamma$ and $Z$ bosons.

 We thus conclude that the essential elements of the profile of
Higgs bosons in the intermediate mass range can be scanned very accurately
at $e^+e^-$ colliders.

\vspace*{2cm}
\noindent {\bf Acknowledgements.} We enjoyed interesting conversations with K.
Hagiwara and M.L. Stong on their discussion of the $VVH$ coupling, which partly
overlaps with our analysis. Helpful comments by J.F. Gunion are also gratefully
acknowledged. This research was supported in part by the University of
Wisconsin Research Committee with funds granted by the Wisconsin Alumni
Research Foundation, in part by the U.S.~Department of Energy under contract
no.~DE-AC02-76ER00881, and in part by the Texas National Laboratory Research
Commission under grant no.~RGFY93-221.

\newpage

 \subsection*{Figure Captions}

\begin{enumerate}

\item[Fig.~1.]
Update of (a) the total decay width and (b) the decay branching fractions
of the Higgs boson in the $\cal SM$.
The top quark mass is assumed to be $m_t = 150$~GeV.

\item[Fig.~2.]
Feynman diagram for the bremsstrahl production process $e^+e^- \rightarrow
HZ$.

\item[Fig.~3.]
Total cross sections for the bremsstrahl process: (a) as a function of the
Higgs mass and for three energy values $\sqrt{s}=300,400$ and 500 GeV
and (b) as a function of the center-of-mass energy for three values of
the Higgs mass $M_H=120, 150$ and 180 GeV. While the dashed curves represent
the cross sections in the Born approximation, the solid curves include the
one--loop electroweak corrections assuming $m_t=150$~GeV.

\item[Fig.~4.]
Beamstrahl corrections to the $e^+e^- \rightarrow  ZH$ total cross sections as
a function of the Higgs mass and for three energy values $\sqrt{s}=300,\ 400$
and 500 GeV.

\item[Fig.~5.]
The missing mass distributions for the $e^+e^- \rightarrow  \mu^+ \mu^- H$
signal as a function of the recoil mass in the three Higgs decay channels (a)
$H\rightarrow b\bar{b}$ for $M_H=100, 125$ and 150 GeV, (b) $H \rightarrow
WW^{(*)}$ for $M_H= 120, 150$ and 180 GeV and (c) $H \rightarrow  ZZ^{(*)}$
for $M_H= 150, 200$ and 250~GeV. The c.m.~energy is fixed to $\sqrt{s}=500$~GeV
and an angular cut $|\cos \theta_{\mu \mu}| <0.6$ of the $\mu^+\mu^-$ pair
momentum with respect to the $e^\pm$ axis has been used; the effects of
bremsstrahlung, beamstrahlung and the beam-energy spread are taken into
account.

\item[Fig.~6.]
Same as in Fig.~5 but for $\sqrt{s}=300$ GeV. Note that there
is no $e^+e^- \rightarrow  t\bar{t}$ background.

\item[Fig.~7.]
Angular distributions for the processes $e^+e^- \rightarrow  ZH$, $e^+e^-
\rightarrow  ZA$ and $e^+e^- \rightarrow  ZZ$ for a Higgs mass of 120 GeV and
 a c.m.\ energy of $\sqrt{s}=500$ GeV.

\item[Fig.~8.]
Cross sections for the production of longitudinal vector bosons normalized to
the total cross sections for the processes $e^+e^- \rightarrow  ZH$ as a
function of the c.m.\ energy for $M_H= 150$ GeV.

\item[Fig.~9.]
Definition of the polar and azimuthal angles (a) in the production process
$e^+e^- \rightarrow  HZ \ [Z \rightarrow  f\bar{f}]$, and (b) in the decay
processes $H \rightarrow  VV^{(*)} \rightarrow  (f_1\bar{f_2})(f_3\bar{f_4})$.

\item[Fig.~10.]
The width of Higgs decays to longitudinally polarized vector bosons normalized
to the total width for the decays $H \rightarrow  VV^{(*)}$ as a function of
the mass ratio $M_{H}/(2M_V)$.

\item[Fig.~11.]
The distributions with respect to the invariant mass of the off--shell vector
bosons in the decay processes $H \rightarrow  ZZ^*$ and $A \rightarrow  ZZ^*$
for $M_H=M_A=150$~GeV.

\end{enumerate}
\end{document}